\documentclass[doublecol,linenumbers]{epl2}
\usepackage{amssymb,amsmath}
\usepackage{indentfirst}
\usepackage{color}
\newcommand{\be}{\begin{equation}}
\newcommand{\ee}{\end{equation}}

\newcommand{\ba}{\begin{eqnarray}}
\newcommand{\ea}{\end{eqnarray}}

\title{Classical Tests of General Relativity: Brane-World Sun from Minimal Geometric Deformation}
\shorttitle{Classical Tests of General Relativity: Brane-World Sun from Minimal Geometric Deformation}
\author{R. Casadio\inst{1,2} \and  J. Ovalle\inst{1,3} \and Rold\~ao da Rocha\inst{4}}
\shortauthor{R. Casadio \etal}
\institute{      
\inst{1}
Dipartimento di Fisica e Astronomia, Universit\`a di Bologna, 
via Irnerio 46, 40126 Bologna, Italy
\\              
\inst{2}
INFN, Sezione di Bologna, viale B.~Pichat 6, 40127 Bologna, Italy
\\
\inst{3}
Departamento de F\'isica, Universidad Sim\'on Bol\'ivar,
Apartado 89000, Caracas 1080A, Venezuela
\\
\inst{4} CMCC,
Universidade Federal do ABC, 09210-170, Santo Andr\'e, SP,
Brazil}
\abstract{We consider a solution of the effective four-dimensional brane-world equations,
obtained from the General Relativistic Schwarzschild metric via the principle of Minimal
Geometric Deformation, and investigate the corresponding signatures stemming from the
possible existence of a warped extra-dimension. 
In particular, we derive bounds on an extra-dimensional parameter, closely related with
the fundamental gravitational length, from the experimental results of the classical tests
of General Relativity in the Solar system.}
\begin{document}
\maketitle
\section{Introduction}
Brane-world (BW) models~\cite{r1} represent a well-known branch of contemporary
high-energy physics, inspired and supported by string theory.
These models are indeed a straightforward 5D phenomenological realisation
of the Ho$\check{\rm r}$ava-Witten supergravity solutions~\cite{HW}, when the hidden brane is
moved to infinity along one extra-dimension, and the moduli effects from the remaining compact
extra-dimensions may be neglected~\cite{maartens}.
The brane self-gravity, encoded in the brane tension $\sigma$, is one of the fundamental
parameters appearing in all BW models, with $\sigma^{-1/2}$ playing the role of
the (5D) fundamental gravitational length scale~\footnote{We shall mostly use
units with the four-dimensional Newton constant $G=c=1$, unless otherwise specified.}.
In this work, we shall explicitly study the observational effects determined by the
parameter $\beta\simeq (\sigma^{-1/2}/R)^2$, which describes a candidate for the 
modified 4D geometry surrounding a star of radius $R$in the BW.
{This particular geometry will be obtained as an exact minimal geometric
deformation (MGD)~\cite{jovalle2009} of the Schwarzschild solution to the field
equations in General Relativity (GR).
The MGD approach ensures, by construction, that this BW solution smoothly reduces to
the GR Schwarzschild metric in the limit $\sigma^{-1}\to 0$,
thus allowing us to analyse variations from GR predictions for small values of the
deforming parameter $\beta$.
This parameter controls the corrections and is related to the brane tension,
the radius of the star, and will also be shown to depend on the compactness of the star.} 
\section{Minimal Geometric Deformation}
\label{MGD}
The effective Einstein equations on the brane take the form~\cite{GCGR}
\begin{equation}
\label{gmunu}
G_{\mu\nu}=
-\tilde{T}_{\mu\nu}-\Lambda\, g_{\mu\nu}
\ ,
\end{equation}
{}{where
$\tilde{T}_{\mu\nu}
=
T_{\mu\nu}+\frac{6}{\sigma}\,S_{\mu\nu}
+\frac{1}{8\,\pi}\,{\cal E}_{\mu\nu}
$
denotes the effective energy-momentum tensor,}
with $T_{\mu\nu}$ the stress tensor of brane matter, and
$\cal{E}_{\mu\nu}$ and $S_{\mu\nu}$ the non-local and high-energy
Kaluza-Klein corrections. 
If BW matter is a perfect fluid with 4-velocity $u^\mu$,
and $h_{\mu\nu}=g_{\mu\nu}-u_\mu u_\nu$ are the components of the metric tensor
orthogonal to the fluid lines, then
\begin{equation}
\label{emunu}
{\cal E}_{\mu\nu}
=
\frac{6}{\sigma}\left[{\cal U}\left(u_\mu\,u_\nu+\frac{1}{3}\,h_{\mu\nu}\right)
+{\cal P}_{\mu\nu}\!+\!{\cal Q}_{(\mu}\,u_{\nu)}\right]
\ ,
\end{equation}
where ${\cal U}$ denotes the bulk Weyl scalar,  ${\cal P}_{\mu\nu}$ is
the anisotropic stress and ${\cal Q}_\mu $ the energy flux.
\par
Solving the effective 4D  Einstein equations in the BW is a hard task and,
already in the simple case of a spherically symmetric metric,
{}{
\begin{equation} 
ds^2
=
e^\nu\,dt^2-e^\lambda\,dr^2-r^2d\Omega^2
\ ,
\label{gsph}
\end{equation}
only a few ``vacuum'' solutions are known
analytically~\cite{wiseman,dejan,dadhich,page,cfabbri}.}
Moreover, for static matter distributions $Q_\mu =0$ and
${\cal P}_{\mu\nu}={\cal P}\left(r_\mu\, r_\nu+\frac{1}{3}\,h_{\mu\nu}\right)$~\cite{maartens}. 
For stellar systems, the quest for BW solutions becomes even more intricate,
mainly due to the presence of extra terms, non-linear in the matter fields,
which emerge from high-energy corrections~\cite{maartens,GCGR,aliev}.
Nonetheless, two approximate analytical solutions have been found in the MGD approach.
It is worth to emphasize that these metrics are exact solutions of the effective
equations~\eqref{gmunu}, although they are not complete solutions of the full 5D
equations~\cite{ovalle2007,tolman}.
This approach also yields physically acceptable interior solutions for stars~\cite{jovalleBWstars},
relates the exterior tidal charge found in Ref.~\cite{dadhich} to the ADM mass, and
let us study (micro) black holes~\cite{covalle1, covalle2},
elucidates the role of exterior Weyl stresses from bulk gravitons on compact stellar
distributions~\cite{olps2013} and shows the existence of BW stars with Schwarzschild
exterior without energy leaking into the bulk~\cite{darkstars}.
Moreover, both the associated 5D solutions and black strings were obtained in various
contexts~\cite{bs1,bs22,bas22,PRD,Casadio:2013uma,Bazeia:2014tua},
with models for the quasar luminosity variation induced by BW
effects~\cite{daRocha:2013ki,CoimbraAraujo:2005es}.
\par
Let us start by revisiting the MGD approach, which is built
on the requirement that GR must be recovered in the low energy limit
$\sigma^{-1}\to 0$.
In particular, by solving the effective 4D equations~\eqref{gmunu},
the radial component of the metric is deformed by bulk effects and can be written
as~\cite{covalle2}
\begin{eqnarray}
\label{edlrwssg}
e^{-\lambda}
=
\mu+f
\ ,
\end{eqnarray}
{}{where 
\begin{equation}
f
=
e^{-I}\left(\beta
+
\int_{r_0}^r\!\!\frac{e^I\,d x}{\frac{\nu'}{2}\!+\!\frac{2}{x}}
\left[H+\!\frac{1}{\sigma}\left(\rho^2\!+\!3\,\rho p\right)\right]
\right)
\ ,
\label{f(r)}
\end{equation}
\begin{equation}
I(r,r_0)
\equiv
\int^r_{r_0}
\frac{\nu''+\frac{{\nu'}^2}{2}+\frac{2\nu'}{x}+\frac{2}{x^2}}
{\frac{\nu'}{2}+\frac{2}{x}}\,dx
\ ,
\label{I}
\end{equation}
and
\begin{eqnarray}
\mu
\!\!&\!\!=\!\!&\!\!
\begin{cases}
\label{intmass}
1-\strut\displaystyle\frac{2\,M}{r}
\ ,
&
\mbox{for}\, r>R
\ ,
\\
1-\strut\displaystyle\frac{8\,\pi}{r}\!\!\int_0^r\! x^2\rho\, dx
\equiv
1-\frac{2\,m(r)}{r}
\,,
&
\mbox{for}\, r\leq R
\ ,
\end{cases}
\nonumber
\end{eqnarray}
where $m$ denotes the standard GR interior mass function.}
The constant $M$ depends in general on the brane tension
$\sigma$ and must take the value of the GR mass $M_0 = m(R)$
in the absence of BW effects,
namely
$M_0 = M\vert_{\sigma^{-1}=0}$. 
The function $H$ in~\eqref{edlrwssg} is given by
\begin{eqnarray}
\label{H}
H(p,\rho ,\nu )
&\!\!\equiv\!\!&
24\,\pi\,p-\left[ \mu ^{\prime }\left( \frac{\nu
^{\prime }}{2}+\frac{1}{r}\right)\right.
\nonumber
\\
&&
\left.+\mu \left( \nu ^{\prime \prime }
+\frac{\nu ^{\prime 2}}{2}\!+\!\frac{2\nu ^{\prime }}{r}
+\frac{1}{r^{2}}\right)
\!-\!\frac{1}{r^{2}}\right]
\ ,
\end{eqnarray}
and encodes anisotropic effects due to the bulk gravity on the pressure $p$,
matter density $\rho$ and the metric function $\nu$.
Finally, the parameter $\beta$ in~\eqref{f(r)} depends 
on the brane tension $\sigma$, the radius $R$ and the mass $M$
of the self-gravitating system, and must be zero in the GR limit.
In the interior, $r<R$, the condition $\beta=\beta(\sigma,R,M)=0$ must hold
in order to avoid singular solutions at $r=0$ [since the integral in~\eqref{I} would
diverge for $r_0\to 0$].
However, for a vacuum solution, or more properly, in the region $r>R$ where there is
only a Weyl fluid surrounding the spherically symmetric star, the parameter $\beta$
can differ from zero.
\par
The crucial point is that, any given perfect fluid solution in GR yields
$H(p,\rho ,\nu )=0$, which provides the foundation for the MGD approach.
In fact, every perfect fluid solution in GR can be used to produce a \textit{minimal\/}
deformation on the radial metric component~\eqref{edlrwssg}, in the sense that all the
deforming terms in Eq.~\eqref{f(r)} are removed, except for (a) those produced
by the density $\rho$ and pressure $p$, which are always present
in a realistic stellar interior (where $\beta=0$ for $r<R$),
and (b) the one proportional to the parameter $\beta$ in a vacuum exterior (with
$p=\rho=0$ for $r>R$).
{}{It is worth to emphasise that the condition $H=0$ holds for any BW solution
obtained by the MGD approach, and corresponds to a minimal
deformation in the sense explained above.
Moreover, $H$ may not be negative when a {\it perfect fluid\/} is
used as the gravitational source on the brane, since~\cite{jovalle2009} 
$
H(p,\rho ,\nu )
\equiv
24\,\pi\,p-\left(2\,G^2_{\ 2}+G^1_{\ 1}\right)\mid_{\sigma^{-1}=0}
$
and the components of the Einstein tensor $G^1_{\ 1}=G^2_{\ 2}=8\,\pi\,p$
for a spherically symmetric perfect fluid, so that the condition $H=0$ 
always holds.}
\par
In order to obtain a deformed exterior geometry, we then start by inserting
the spherically symmetric Schwarzschild metric
\be
\label{components}
e^{\nu_S}=e^{-\lambda_S}
=
1-\frac{2\,M}{r}
\ ,
\ee
in the expression~\eqref{edlrwssg} for $r>R$, where $p=\rho=0$.
Since~\eqref{components} is a GR solution, $H(r>R)=0$ and
the correction in Eq.~\eqref{f(r)} will thus be minimal, 
\begin{equation}
\label{def}
f^+(r)
\equiv
\left.f(r)\right|_{p=\rho=H=0}
=
\beta\,e^{-I}
\ .
\end{equation}
The outer radial metric component~\eqref{edlrwssg} will read
\begin{eqnarray}
\label{g11vaccum}
e^{-\lambda^+}
=
{1-\frac{2\,M}{r}}
+\beta(\sigma,R,M)\,e^{-I}
\ ,
\end{eqnarray}
which clearly represents a BW solution different from the GR Schwarzschild
metric, with $\beta$ equal to the extra-dimensional correction to
the GR vacuum evaluated at the star surface, that is $\beta=f^+(r=R)$, and
containing a ``Weyl fluid'' for $r>R$~\cite{olps2013}.
\par
We next consider the general matching conditions between a general interior
MGD metric (for $r<R$),
\begin{equation}
\label{genint}
ds^2
=
e^{\nu^-(r)}\,dt^2-\frac{dr^2}{1-\frac{2m(r)}{r}+f^-(r)}-r^2\,d\Omega^2
\ ,
\end{equation}
where $f^-$ is also given by Eq.~\eqref{f(r)} with $H=0$, 
and the above exterior metric (for $r>R$), which can be written
like~\eqref{genint} by replacing $-$ with $+$. 
Continuity of the metric at the star surface $\Sigma$ of radius $r=R$
yields
\begin{eqnarray}
\label{ffgeneric1}
\nu^-_R
=
\nu^+_R,\qquad
\frac{2\, M}{R}
=
\frac{2\, M_0}{R}
+
\left(f^+_R-f^-_R\right)
\ ,
\end{eqnarray}
where $F_R^\pm\equiv F(r\to R^\pm)$ for any function $F$.
Continuity of the second fundamental form on $\Sigma$ likewise
provides the expression
$
\left[G_{\mu\nu}\,r^\nu\right]_{\Sigma}
=
0$, where $r_\mu$ denotes a unit radial vector and
$[f]_{\Sigma}\equiv f(r\to R^+)-f(r\to R^-)$.
On using the effective 4D equations~\eqref{gmunu}, this condition 
becomes
\be
\label{matching3}
\left[
p+\frac{1}{\sigma}\left(2\,\mathcal{U}+\frac{\rho^2}{2}+\rho\, p
\right)+4\,\frac{\cal{P}}{\sigma}
\right]_{\Sigma}
=0
\ .
\ee
Since the star is assumed to be only surrounded by a Weyl fluid
described by ${\cal U}^+$ and ${\cal P}^+$ (and $p=\rho=0$) for $r>R$,
this matching condition takes the final form
\begin{eqnarray}
\label{match11}
\sigma p_R
\!+\!
4\,{\cal P}_R^-
+2{\cal U}_R^-+\frac{\rho_R^2}{2}
+\rho_Rp_R
\!=\!
{2}\!
\left(
2\,{\cal P}_R^+
+{\cal U}_R^+
\right)
\end{eqnarray}
with $p_R\equiv p_R^-$ and $\rho_R\equiv \rho_R^-$.
The limit $\sigma^{-1}\rightarrow 0$ in Eq.~\eqref{match11} leads to the well-known
GR matching condition $p_R =0$ at the star surface.
Eqs.~\eqref{ffgeneric1} and~\eqref{match11}
are the necessary and sufficient conditions for the matching of the interior MGD metric
to a spherically symmetric ``vacuum'' filled by a BW Weyl fluid~\cite{germ}. 
\section{BW star}
\label{IVA}
BW effects on spherically symmetric stellar systems have already
been extensively studied (see, e.g.~Refs.~\cite{HarkoLake}
for some recent results).
Let us now investigate in details the MGD function $f^+(r)$ produced by
the Schwarzschild solution~\eqref{components}.
By inserting it into Eq.~\eqref{def}, we obtain
\begin{equation}
\label{hhh}
f^+(r)
=
\beta(\sigma,R,M)\,
\frac{b}{r}\,\frac{1-\frac{2M}{r}}{1-\frac{3M}{2\,r}}
\ ,
\end{equation}
where $b$ is a length given by
$b
\equiv
R{(1-\frac{3M}{2R})}/{(1-\frac{2M}{R})}$ 
and the deformed exterior metric components read
\begin{subequations}
\ba
\label{nu}
e^{\nu^+}
&\!\!=\!\!&
1-\frac{2\,M}{r}
\ ,
\\
e^{-\lambda^+}
&\!\!=\!\!&
\left(1-\frac{2\,M}{r}\right)
\left[1+\frac{\beta(\sigma,R,M)}{1-\frac{3\,M}{2\,r}}\,\frac{b}{r}\right]
\ ,
\quad
\label{mu}
\ea
\end{subequations}
matching the vacuum solution found in Ref.~\cite{germ}  
when $\beta\,b={K}/{\sigma}$, with $K>0$. 
The corresponding Weyl fluid is described by~\cite{olps2013}
\ba
\label{pp2}
\frac{{\cal P}^+}{\sigma}
=
\frac{\beta\,b\left(1-\frac{4\,M}{3\,r}\right)}{9\,r^3\left(1-\frac{3\,M}{2\,r}\right)^2}
\ ,
\
\frac{{\cal U}^+}{\sigma}
=
\frac{-\beta\,b\,M}{12\,r^4\left(1-\frac{3\,M}{2\,r}\right)^2}
\ .
\ea
\par
We can now obtain the parameter $\beta=\beta(\sigma,R,M)$,
depending on the interior structure, by employing
the deformed Schwarzschild metric~\eqref{nu} and \eqref{mu}
in the matching conditions~\eqref{ffgeneric1} and~\eqref{match11}.
Eq.~\eqref{ffgeneric1} just becomes
$ 
e^{\nu^-_R}
=
1-\frac{2M}{R}$, whereas Eq.~\eqref{match11} yields
\be
\label{sfgeneric}
p_R
+\frac{f^-_R}{R}
\left(\nu'_R+\frac{1}{R}\right)
=
-\frac{f^+_R}{R^2}
\ , 
\ee
with $\nu'_R\equiv(\nu^-)'\!\!\mid_{r=R}$. 
These are the necessary and sufficient conditions for matching
the two minimally deformed metrics given by Eqs.~\eqref{genint}, \eqref{nu}
and~\eqref{mu}.
If $M$ in Eq.~\eqref{components} were the GR mass $M_0$, one would have
$f_{R}^+=f_{R}^-$ [see Eq.~\eqref{ffgeneric1}],
which is an unphysical condition, according to Eq.~\eqref{sfgeneric}.
In fact, the interior deformation $f=f^-(r)$ is positive, 
but the matching condition~\eqref{sfgeneric} shows that the exterior deformation
must be negative at the star surface, $f^+_R<0$, or else a negative pressure
$p_R < 0$ would appear.
Hence, according to Eq.~\eqref{hhh}, the deformation $f^+(r>R)$
is negative for $\beta<0$~\cite{darkstars}.
\par
The exterior geometry given by Eqs.~\eqref{nu} and \eqref{mu} may seem
to have two horizons, namely
$
r_h = 2\,M $ and $
r_2=3\,M/2 - \beta\,b$. 
However, since $\beta$ must be proportional to $\sigma^{-1}$ in order to 
recover GR, the condition $r_2< r_h$ must hold, and the outer horizon radius
is given by $r_h = 2\,M$.
The specific value $\beta=-M/2$ would produce a
single horizon $r_h = r_2 = 2\,M$, but the limit $\sigma^{-1}\to 0$ 
does not reproduce the Schwarzschild solution, as seen from the
condition $M_0 = M\vert_{\sigma^{-1}=0}$.
On the other hand, $f^+_R < 0$ implies that the deformed
horizon radius $r_h=2\,M$ is smaller than the Schwarzschild radius
$r_H=2\,M_0$, as it can  be clearly realised from Eq.~\eqref{ffgeneric1}.
This general result shows that 5D effects weaken the 
strength of the gravitational field produced by the self-gravitating stellar system.
\par
Finally, when \eqref{hhh} is considered in the
matching condition~\eqref{sfgeneric}, we obtain
\begin{equation}
\label{beta}
\beta
=
f^+_R
=
-R^2
\left[p_R+\left(\frac{1}{R}+{\nu'_R}\right)\frac{f^-_R}{R}\right]
\ ,
\end{equation}
showing that $\beta$ is always negative and (interior) model-dependent
through $\nu'_R$.
In particular, we can find $\beta$ by considering the exact interior BW solution
of Ref.~\cite{ovalle2007}, that is
\begin{eqnarray}
\label{regularmass}
f^-
\!\!\!&\!\!\!=\!\!\!&\!\!
\!\frac{32 C}{49\sigma}
\left[\frac{240\!+\!589Cr^2\!-\!25C^2r^4\!-\!41C^3r^6\!-\!3C^4r^8}{3(1+Cr^2)^4(1+3Cr^2)}\right.\nonumber\\&&\left.
-\frac{80}{(1+Cr^2)^2}\frac{{\rm \arctan}(\sqrt{C}r)}{(1+3Cr^2)\sqrt{C}r}
\right]
\ ,
\end{eqnarray}
where $C$ denotes a constant (with the same dimensions of $\sigma$)
given by $C\,R^2=\frac{\sqrt{57}-7}{2}\equiv\alpha$,
and $\nu'={8Cr}\,({1+Cr^2})^{-1}$. 
Using the explicit form of $f^-(R)$ and $p_R=0$
in Eq.~\eqref{beta} yield
\begin{equation}
\beta(\sigma,R)
=
f_R^+
=
-\frac{C_0}{R^2\,\sigma}
\ .
\label{c0}
\end{equation}
where $C_0\simeq 1.35$ is a (dimensionless) constant. 
The exterior deformation is finally obtained by using Eq.~\eqref{c0}
in Eq.~\eqref{hhh}, leading to
\begin{equation}
\label{hhhh}
f^+
=
-\frac{C_0\,b}{R^2\,\sigma\,r}
\left(
\frac{r-2\,M_0}{r-3\,M_0}
\right)
+{\cal O}(\sigma^{-2})
\ ,
\end{equation}
where $b_0=b(M_0)$ is given by the length $b$ at $M=M_0$.
{}{The deformation $f^+(r>R)$ is therefore a monotonically increasing
function of the star compactness $M_0/R$.
Since extra-dimensional effects are the strongest at the surface
$r=R$ and become more important for smaller stellar distributions,
the more compact the star the larger $\beta$, and thus the MGD
of the GR solution.}
\section{Solar System Classical Tests}
Classical tests in the Solar system can probe BW signatures.
The perihelion precession of Mercury, the deflection of light by the Sun
and the radar echo delay observations are well-known tests for the
Schwarzschild solution of GR and, in BW models, for the DMPR and the
Casadio-Fabbri-Mazzacurati metrics as well.
BW effects in spherically symmetric space-times were comprehensively
studied, e.g., in Ref.~\cite{boemer}.
In our case, Solar system tests will be employed to bound the MGD parameter
$\beta$ in Eq.~\eqref{c0}. 
\subsection{Perihelion Precession}
{}{A test particle in a spherically symmetric metric~\eqref{gsph}
has two constants of motion, $E$ and $L$, respectively yielding energy and
angular momentum conservation.}
By the usual change of variable $r=1/u$ and defining 
\begin{eqnarray}
\label{lamb}
g(u)=1-e^{-\lambda}
\ ,
\end{eqnarray}
the relevant equation of motion reads~\cite{boemer}  
\begin{eqnarray}
\frac{d^{2}u}{d\phi ^{2}}\!+\!u
\!=\!
\frac{1}{2}\frac{d}{du}\!\!\left(\!\frac{E^{2}e^{-\lambda-\nu}}
{c^{2}L^{2}}\!-\!\frac{e^{-\lambda}}{L^{2}}\!+\!g(u)u^{2}\!\right)
\!\equiv k(u)
\ .
\label{fu}
\end{eqnarray}
By denoting $\gamma(u)=\left({1-\left( {dk}/{du}\right)\vert_{u_{0}} }\right)^{1/2}$,
a circular orbit $u=u_{0}$ is determined by the root of the equation
$u_{0}=k(u_{0})$, and a deviation with respect to it is provided by 
$\delta =\delta _{0}\cos \left(\gamma(u)\phi +\alpha \right)$,
with $\delta _{0}$ and $\alpha$ constants~\cite{boemer}.
The variation of the orbital angle with respect to successive perihelia is given by 
$\phi =\frac{2\pi }{\gamma(u)}=\frac{ 2\pi }{1-\iota},$ where the perihelion
advance is  
$\iota \simeq\frac{1}{2}\left( \frac{dk}{du}\right) _{u=u_{0}}$,
for small values of $\left(dk/du\right) _{u=u_{0}}$.
For a complete rotation, the perihelion advance is $\delta \phi \approxeq 2\pi \iota $. 
\par
We now consider the perihelion precession of a planet in the MGD geometry
described by Eqs.~(\ref{nu}) and (\ref{mu}).
Since $L$ is related to the orbit parameters by $L=2\pi a^{2}\sqrt{1-e^{2}} /cT$~\cite{boemer},
where $T$ denotes the period of motion, the perihelion advance thus  yields
\be
\delta \phi \!=\!\delta \phi _{GR}-{f}(\beta)
\ ,
\label{ppp}
\ee
where $\delta \phi _{GR}=6\pi GM/c^{2}a\left(1-e^{2}\right)$ is the well-known
Schwarzschild precession formula and $
{f}(\beta)
\simeq
673.94\,\beta$, where we employed 
$c=2.998\times 10^{8}~{\rm m/s}$,
$M_{\odot}=1.989\times 10^{30}~{\rm kg}$, 
$a=57.91\times 10^{9}~{\rm m},$
$R_{\odot}=6.955 \times 10^8~ {\rm m}$,
$e=~0.205615$, and $G=6.67\times 10^{-11}~{\rm m^{3}kg^{-1}s^{-2}}$. 
\par
The observed difference
$\delta{\phi}-\delta \phi _{GR}=0.13\pm 0.21\,$arcsec/century~\cite{Sh} 
can thus be ascribed to BW effects, according to Eq.~\eqref{ppp}.  
Observational data~\cite{Sh,boemer} yield the bound
$
f(\beta) \leq (1.89 \pm 2.33) \times 10^{-8}$, 
which constrains 
\begin{equation}
\beta  \lesssim (2.80 \pm 3.45) \times 10^{-11}\ .
\label{betal}
\end{equation}
\subsection{Light Deflection}
A similar procedure describes photons on null geodesics,
with the equation of motion that can be written as
\begin{equation}
\left( \frac{du}{d\phi }\right) ^{2}+u^{2}
=
\frac{1}{c^{2}}\frac{E^{2}}{L^{2}}\,e^{-\nu -\lambda}
+g(u)\,u^{2}
\equiv p(u)
\ ,
\label{puu}
\end{equation}
which therefore implies
 $\frac{d^{2}u}{d\phi ^{2}}+u=\frac{1}{2}\frac{dp(u)}{du}$.
In the lowest approximation, the solution is 
$u=\frac{\cos \phi }{R_0},$ where $R_0$ is the distance of
closest approach to the mass $M$.
It can be iteratively employed in the above equation, yielding
${d^{2}u}/{d\phi ^{2}}+u
=
\frac{1}{2}{d}\left[p\left( \frac{\cos \phi }{R_0}\right)\right]/du$. 
The total deflection angle of the light ray is given by
$\delta =2\varepsilon $ \cite{boemer}.
\par
For the geometry~(\ref{nu}) and (\ref{mu}), Eq.~(\ref{lamb}) leads to
$g(u)=\left(2GM/c^{2}\right) u$, resulting in
\begin{eqnarray}
p(u)
\!\!&\!\!=\!\!&\!\!
\frac{\beta\,b_0}{\left(2- \frac{3GMu}{c^2}\right)^2}
\left\{ u^2 \left[{\frac{GMu}{c^2}
\left(\frac{9GM}{c^2} u\!-\!11\right)\!+\!3}\right]
\right.
\nonumber
\\
&&
\left.
\phantom{\frac{\beta\,b_0}{\left(2- {3GMu}\right)^2}}
-{2 a^2}\right\}+\frac{3GMu^2}{c^2}
\ .
\label{pudef}
\end{eqnarray}
The total deflection of light is finally given by 
\be
\delta \phi
=
\frac{4GM}{c^2R_0}
+\beta\,b_0 \left(\frac{E^2\, R_0}{c^2\,L^2}+\frac{18 \pi  c^2\,R_0}{G\,M}\right)
\ ,
\label{a1}
\ee
in the limit $\left(\frac{GM}{c^2\,R_0}\right)^2\ll 1$, $\frac{M}{L}\ll 1$ and
$\frac{E^2}{c^2}-1\ll1$, which implies the bound
\be
\beta\lesssim (1.07 \pm 4.28) \times 10^{-10}
\ .
\ee
\subsection{Radar Echo Delay}
Another classical test of GR measures the time for radar signals to travel to,
for instance, a planet~\cite{Sh}.
The time for light to travel between two planets, respectively at a distance
$\ell_{1}$ and $\ell_{2}$ from the Sun, is well-known to be
${\rm T}_{0}=\int_{-\ell_{1}}^{\ell_{2}}dx/c$.
On the other hand, if light travels in the vicinity of the Sun, the time lapse
$\delta {\rm T = T-T}_0$ is given by~\cite{boemer} 
\begin{equation}
\label{delay}
\delta {\rm T}
=
\frac{1}{c}\int_{-\ell_{1}}^{\ell_{2}}\!\!\left\{e^{{\left[\lambda \left( \sqrt{ x^{2}+R^{2}}\right)
-\nu \left( \sqrt{x^{2}+R^{2}}\right) \right]/2}}-1\right\}
dx
\ ,
\end{equation}
since $r=\sqrt{x^{2}+R^{2}}$.
The above integrand takes the form 
\begin{eqnarray}
&
\exp\! \left(\frac{\lambda-\nu}{2}\right)\!-1
=
\left({1\!-\!\frac{2 G M}{c^2r}}\right)^{\!-1} \!\!
\left(\frac{2 \beta \ell_0 }{3 GM\!-\!2 c^2r}\!+\!1\right)^{-\frac12}
&
\nonumber
\\
&
\approx
\strut\displaystyle{
\frac{2 GM}{c^2r}
-\frac{ \beta \ell_0 }{3 GM\!-\!2c^2r}\!+\!\frac{4 \beta \ell_0  GM}{c^2r (3 GM\!-\!2 c^2r)}
\ ,}
&
\end{eqnarray}
where we used a first order approximation based on Eqs.~(\ref{nu}) and (\ref{mu}). 
Eq.~(\ref{delay}), using the approximations $R^{2}/\ell_{i}^{2}\ll 1$ ($i=1,2$), and considering terms
up to order $(GM/c^2R)^2$, hence reads
\begin{equation}
\delta {\rm T}  
\simeq
\delta {\rm T_{GR}}
+\frac{\beta \ell_0}{c^3R}
\left[\ln\! \left(\!\frac{4\ell_{1}\ell_{2}}{R^{2}}\!\right)\!-\!\frac{5\pi GM}{2}\right]
\ ,
\label{good}
\end{equation}
which reproduces the Schwarzschild radar delay
$\delta {\rm T_{GR}}= \frac{2GM}{c^{3}}\ln \frac{4\ell_{1}\ell_{2}}{R^{2}}$
when $\beta=0$, and the second term imposes a constraint on BW models.
Recent measurements of the frequency shift of radio photons both 
to and from the Cassini spacecraft, as they passed near the Sun, 
have refined the observational constraints on the radio
echo delay.
For the time delay of the signals emitted on Earth towards the Sun, one obtains
$\Delta t_{\rm radar}=\Delta t_{\rm radar}^{\rm GR}\left( 1+\Delta _{\rm radar}\right) $, with
$\Delta_{\rm radar}\simeq (1.1 \pm 1.2) \times 10^{-5}$~\cite{Reasemberg}.
In the BW geometry~\eqref{nu} and \eqref{mu},
measurements of the frequency shift of radio photons~\cite{boemer,Reasemberg}
finally yield the physical bound
\begin{equation}
\beta
\lesssim
\frac{5\pi G^2M^2\Delta_{\rm radar}}{2\,\ell_0\,R\ln\left(\frac{4\ell_1\ell_2}{R^2}\right)}
\simeq
(3.96 \pm 4.30)
\times 10^{-5}
\ .
\end{equation}
This provides a bound on the MGD parameter $\beta$, which is the weakest one among
those in our analysis.
\section{Concluding Remarks}
BW models can be confronted with astronomical and
astrophysical observations at the Solar system scale.
In this paper we have in particular considered the BW exterior
solution~\eqref{nu} and~\eqref{mu} obtained by means of the MGD procedure,
and compared its predictions with standard GR results.
This exterior geometric contains a parameter $\beta$ and
we were able to constrain it from the presently available observational
data in the Solar system. 
We found the strongest constraint is given by measurements of the perihelion
precession, namely Eq.~\eqref{betal}.
\par
{}{
Let us recall that limits for the brane tension in the DMPR and Casadio-Fabbri-Mazzacurati
BW solutions have already been determined via the classical tests of GR~\cite{boemer}.
Since bounds on the parameter $\beta$ imply lower bounds for the brane tension
from Eq.~(\ref{c0}), we can conclude that the constraint~(\ref{betal}) complies with
the ones provided by such solutions of the effective 4D Einstein equations~\eqref{gmunu}.
In fact, the brane tension in the MGD framework is bounded according to
\begin{equation}
\sigma
\geq
\frac{9M_\odot c^2}{\pi R_\odot^3\beta }\,
\frac{\left(1-\frac{2GM_\odot}{c^2 R_\odot}\right)^2}{\left(1-\frac{3GM_\odot}{2c^2 R_\odot}\right)}
\ ,
\end{equation}
which implies that $\sigma \geq  5.19\times10^6 \;{\rm MeV^4}$, when the bound (\ref{betal})
is taken into account (we omit errors here since we are solely interested in orders of magnitude).
This bound is still much stronger than the cosmological nucleosynthesis constraint,
however much weaker than the lower bound obtained from measurements of the Newton law
at short scales.
We can therefore conclude that the MGD geometry~\eqref{nu} and~\eqref{mu} is
acceptable within the present measurements of BW high-energy corrections.
\par
After this work had been completed, we developed an extension of the MGD,
which produces a rich but complex set of new exterior solutions~\cite{MGDextended}, 
whose complete analysis is highly non-trivial.
However, the solution used here represents the simplest non-trivial extension of
the Schwarzschild solution within the extended MGD approach.}
\vspace*{-0.2cm}
\section{Acknowledgments}
RC is partly supported by INFN grant FLAG.
RdR is grateful to CNPq grants No.~303027/2012-6 and  No.~473326/2013-2
for partial financial support.
JO is partially supported by Erasmus Mundus program,
grant 2012-2646 / 001-001-EMA2.
\end{document}